\documentclass[aps,prc,preprint,superscriptaddress]{revtex4-1}

\usepackage{amsmath,amssymb}
\usepackage{graphicx,float}
\usepackage[utf8]{inputenc}

\DeclareMathAlphabet{\mathpzc}{OT1}{pzc}{m}{it}

\begin{document}


\title{Dynamics of one-dimensional correlated nuclear systems within non-equilibrium Green's function theory}



\author{Hao Lin}
\affiliation{National Superconducting Cyclotron Laboratory and Department of Physics and Astronomy,
Michigan State University, East Lansing, Michigan 48824, USA}

\author{Hossein Mahzoon}

\affiliation{National Superconducting Cyclotron Laboratory and Department of Physics and Astronomy,
Michigan State University, East Lansing, Michigan 48824, USA}

\affiliation{Department of Physics, Washington University in St.\ Louis, St.\ Louis, MO 63130, USA }

\author{Arnau Rios}
\affiliation{University of Surrey, Guildford, Surrey GU2 7XH, United Kingdom}

\author{Pawel Danielewicz}

\affiliation{National Superconducting Cyclotron Laboratory and Department of Physics and Astronomy,
Michigan State University, East Lansing, Michigan 48824, USA}



\begin{abstract}
Theory of non-equilibrium Green's function (NGF) provides a practical framework for studying quantum many-body systems out of equilibrium. Extending the previous mean field approach developed for nuclear systems in one dimension with NGF, we introduce isospin degrees of freedom to the Green's functions and incorporate short-range two-body interactions in the second-order self-consistent approximation to correlations, which represents the scattering of momentum orbitals in the Born approximation. We discuss the preparation of a finite nuclear system and examine the impact of correlations on the ground state. We also excite a finite symmetric nuclear system to oscillate in an isovector dipole mode and explore the dissipation effects in the oscillation. Finally, we demonstrate how to boost a slab to a constant and stable motion in a box, based on Galilean covariance of the theory. The studies in this paper lay the ground for the future exploration of collisions of correlated nuclear systems in one dimension.
\end{abstract}


\maketitle

\section{Introduction}
Quantum mechanics has taken root in a bulk part of contemporary nuclear theory, from the Hartree-Fock method \cite{thouless1960,mang1975,gogny1986} to chiral effective field theory \cite{ordonez1994,machleidt2005towards,machleidt2011chiral} in nuclear structure, from the distorted-wave Born approximation \cite{tamura1971distorted,lee196440} and to the coupled-channel techniques \cite{tamura1969coupled,thompson1988coupled,austern1987continuum} in low-energy nuclear reactions, just to name a few. The development of time-dependent quantum approaches is not on par with their static counterparts, with the most notable one being the time-dependent Hartree-Fock (TDHF) theory \cite{engel1975time,cambiaggio1984time,umar2006three,avez2008pairing,ebata2010canonical,maruhn2014tdhf}. However, the bare form of the TDHF theory is a mean field theory neglecting many-body correlations, which does not provide sufficient stopping and thermalization to describe intermediate or high energy heavy-ion collisions \cite{Loebl2012}. The community of intermediate-energy heavy-ion collisions continues to rely predominantly on semi-classical transport models based on the Boltzmann-Uehling-Uhlenbeck (BUU) equation, where quantum effects such as Pauli-blocking \cite{aichelin1985numerical,PhysRevC.97.034625}, the uncertainty principle, through the use of wavepackets \cite{aichelin1986quantum,aichelin1991quantum}, and antisymmetrization \cite{kanada1995structure,ono2004antisymmetrized}, can be partially incorporated. While a vast majority of the semi-classical transport models have demonstrated their applicability and strengths through comparison with experimental data~\cite{wada2000reaction,borderie2001evidence,danielewicz2000determination,hong2014subthreshold}, questions concerning the validity of the quasi-particle picture and the effects of quantum correlations linger. Moreover, as already pointed out in Ref.~\cite{rios2011towards}, the semi-classical methods, detached from the rapidly developing quantum techniques employed in other fields of nuclear physics, are difficult to improve systematically. Even attempts to take into account the missing classical many-body correlations through the incorporation of fluctuations \cite{suraud1990boltzmann,colonna1998fluctuations,napolitani2013bifurcations,lin2019one} will not be able to recover all effects of quantum correlations, as they are of fundamentally different nature. A fully quantum-mechanical transport model for heavy-ion collisions is called for.

Solving the many-body Schr\"{o}dinger equation exactly is almost always a forbiddingly daunting task. Fortunately, more often than not, one is interested in one-body observables, for which a complete many-body wavefunction is not necessary. One can hopefully obtain enough information from one-body reductions, such as one-body density matrix or one-body Green's function, by integrating out the irrelevant degrees of freedom. In obtaining a closed equation of motion, the TDHF theory approximates the two-body density matrix by products of one-body density matrix and consequently neglects all two-body correlations \cite{engel1975time,RingSchuck}. In contrast, in the non-equilibrium Green's function (NGF) theory, two-body correlations are encapsulated in the self-energy and a systematic order-by-order approximation of the self-energy is possible \cite{rios2011towards,danielewicz1984quantum}. As a result, the NGF theory is more versatile and promising than the TDHF theory in the description of intermediate-energy heavy-ion collisions, where effects of two-body correlations are usually non-negligible.

An NGF code for the description of nuclear collisions in one dimension was developed by some of us, and the mean field dynamics within NGF was discussed in Ref.~\cite{rios2011towards}. The current paper serves as an extension to the previous work. Before we present the extensions and discuss the new results, let us briefly go over some relevant concepts and definitions in the NGF theory. (See Refs.~\cite{rios2011towards,danielewicz1984quantum} for more details.) Within the NGF theory, fermionic many-body systems can be characterized in terms of the single-particle Green's functions
\begin{align}
G^{<}(x_1,t_1;x_2,t_2) &=  i \langle \phi^\dagger(x_2,t_2) \, \phi(x_1,t_1)\rangle \, , \\[.5ex]
G^{>}(x_1,t_1;x_2,t_2) &=  -i \langle \phi(x_1,t_1) \, \phi^\dagger(x_2,t_2)\rangle \, .
\end{align}
Here $\phi$ and $\phi^\dagger$ are annihilation and creation operators in the Heisenberg picture, and the expectation value is taken with respect to the total wavefunction also in the Heisenberg picture. When $t_1 = t_2$, the same-time lesser Green's function $G^{<}$ is equivalent to the one-body density matrix up to a constant imaginary factor~$i$. Thus, the lesser and greater Green's functions are often associated with the particle density and the hole density, respectively. One may also define Green's functions as functions of momentum and time arguments, by replacing the annihilation and creation operators in coordinate space by those in momentum space. Green's functions in coordinate space and Green's functions in momentum space can be converted into one another through Fourier transformation.

The equations of motion for the lesser and greater Green's functions are the so-called Kadanoff-Baym equations \cite{kadanoff1962quantum},
\begin{align}
\label{kbe}
\bigg[i\hbar \, \frac{\partial}{\partial t_1}+\frac{\hbar^2}{2m}\frac{\partial^2}{\partial \mathbf{x}_1^2}\bigg] \, G^\lessgtr(\mathbf{1},\mathbf{2}) = &\int d\mathbf{x}_3 \, \Sigma_\textrm{HF}(\mathbf{1},\mathbf{3})\, G^\lessgtr(\mathbf{3},\mathbf{2}) \nonumber\\
& +\int_{t_0}^{t_1}d\mathbf{3} \, \Sigma^{+}(\mathbf{1},\mathbf{3}) \, G^\lessgtr(\mathbf{3},\mathbf{2}) + \int_{t_0}^{t_2}d\mathbf{3} \, \Sigma^{\lessgtr}(\mathbf{1},\mathbf{3}) \, G^{-}(\mathbf{3},\mathbf{2}) \, , \\
\bigg[-i\hbar \, \frac{\partial}{\partial t_2}+\frac{\hbar^2}{2m}\frac{\partial^2}{\partial \mathbf{x}_2^2}\bigg] \, G^\lessgtr(\mathbf{1},\mathbf{2}) = &\int d\mathbf{x}_3 \, G^\lessgtr(\mathbf{1},\mathbf{3}) \,  \Sigma_\textrm{HF}(\mathbf{3},\mathbf{2})\nonumber\\
& +\int_{t_0}^{t_1}d\mathbf{3} \, G^{+}(\mathbf{1},\mathbf{3}) \, \Sigma^\lessgtr(\mathbf{3},\mathbf{2}) + \int_{t_0}^{t_2}d\mathbf{3} \, G^{\lessgtr}(\mathbf{1},\mathbf{3}) \, \Sigma^{-}(\mathbf{3},\mathbf{2}) \, ,
\end{align}
where we have used shorthand notations such as $\mathbf{1}$ to represent $(\mathbf{x}_1,t_1)$. The different $\Sigma(\mathbf{1},\mathbf{2})$ are the so-called self-energy, and provide an approximate description of correlations in the system. They can be expanded diagrammatically, and will be discussed further in the next section. 
The retarded (+) and advanced ($-$) functions are defined through
\begin{equation}
F^\pm(\mathbf{1},\mathbf{2}) = F^\delta(\mathbf{1},\mathbf{2})\pm\theta[\pm (t_1-t_2)] \, [F^{>}(\mathbf{1},\mathbf{2}) - F^{<}(\mathbf{1},\mathbf{2}) ] \, ,
\end{equation}
where $F^\delta$ represents a singular part at $t_1 = t_2$. It has been shown that one can derive the BUU equation from the Kadanoff-Baym equation by approximating the interacting Green's functions with the free Green's functions in a uniform system \cite{danielewicz1984quantum}, which demonstrates a strong link between the NGF approach and the semi-classical models based on the BUU equation. On the other hand, in contrast to classical transport theories, the correlation integrals, integrating over the entire history, lead to memory effects.

To advance the application of these techniques to nuclear physics, we draw on rigorous NGF results and combine them with phenomenology taking into account the context of numerical limitations and the expectations regarding physical characteristics of the nuclear systems. The aim of this paper is to show the applicability of NGF in a variety of settings that are relevant for nuclear theory, in preparation for the application to simulations of nuclear collisions. The organization is as follows.  In~Sec.~II, we will discuss the introduction of isospin degrees of freedom and the incorporation of short-range two-body interactions. In~Sec.~III, we will describe the preparation of finite correlated nuclear systems within the NGF approach. Sec.~IV examines the isovector oscillation of slabs in one dimension and Sec.~V demonstrates how to boost a finite system to move at a constant velocity. In the end, we will give a summary and remark on the prospects in Sec.~VI.

\section{Extending the non-equilibrium Green's function approach for nuclear systems}\label{II}

To describe realistic nuclear systems at moderate energies, an obvious but nonetheless crucial extension to the previous NGF model \cite{rios2011towards} is to introduce isospin dependence in the Green's functions. This can be readily achieved by introducing two versions of the Green's functions, of the self-energies, and of other relevant quantities, one for the neutron subsystem and the other for the proton subsystem. The separation is rather straightforward in practice, but comes with the cost of doubling the memory requirement for the computation and doubling the number of Kadanoff-Baym equations to be solved simultaneously. The mapping from 1D neutron or proton densities to 3D densities continues~\cite{rios2011towards} to follow the relation from the Hugenholtz-van Hove theorem: $\rho^{3D}_{n/p} = \xi \, \rho^{1D}_{n/p}$ with $\xi \approx 0.1217\,$fm$^{-2}$. In~what follows, the subscripts $n$ and $p$, indicating the particle species, will be dropped for the sake of brevity whenever no confusion arises.

The bulk part of mean-field interaction $U$ independent of isospin follows closely from that in Ref. \cite{rios2011towards},
\begin{equation}
U(\rho) = \frac{3}{4}t_0\rho+\frac{2+\sigma}{16}t_3\rho^{1+d},
\end{equation}
with the parametres $t_0 = -2150.1 \textrm{MeV}\,\textrm{fm}^3$, $t_3 =14562 \textrm{MeV}\,\textrm{fm}^{3(1+d)}$ and $d = 0.257$ fitted to reproduce properties of symmetric nuclear matter at saturation density. With the mean-field $U$ dependent on the total baryon density,  the evolutions of the neutron Green's functions and the proton Green's functions are coupled right from the very beginning. Upon differentiating the isospin degrees of freedom, we also introduce a dependence on isospin imbalance, $\delta = (\rho_n-\rho_p)/\rho$, into the nuclear equation of state where the form valid up to the second order in the imbalance is $\frac{E}{A}(\rho, \delta) = \frac{E}{A}(\rho, 0) + S(\rho) \, \delta^2$, with $S(\rho) = S_\textrm{kin}(\rho) + S_\textrm{int}(\rho)$. The first term $S_\textrm{kin}(\rho)$ stems from the difference in kinetic energies of the neutrons and protons for finite imbalance, while the true isospin dependence of the nuclear forces is reflected in the second term taken here in the form
\begin{equation}
S_\textrm{int}(\rho) = S_0 \, \bigg(\frac{\rho}{\rho_0}\bigg)^{\sigma} \, ,
\label{eq:Sint}
\end{equation}
where the values $S_0=20.1\,$MeV, $\rho_0=0.16\,$fm$^{-3}$ and $\sigma=0.35$ are currently used in the model. The isospin-dependent component of the mean field potential $U_{n/p}^\textrm{sym}(\rho,\delta)$ is given by
\begin{equation} 
U_{n/p}^\textrm{sym}(\rho,\delta) = \partial[\rho \, S_\textrm{int}(\rho) \, \delta^2] \Big/ \partial\rho_{n/p}.
\end{equation} 
The Hartree-Fock self-energy $\Sigma_\textrm{HF}$ is approximated as a sum of an isospin-independent component and an isospin dependent component,
\begin{equation}
\label{hf}
\Sigma_\textrm{HF}(\mathbf{1},\mathbf{2}) = \delta(\mathbf{1}-\mathbf{2})\{U[\rho(\mathbf{1})]+U_{n/p}^\textrm{sym}[\rho(\mathbf{1}),\delta(\mathbf{1})]\}.
\end{equation}
The introduction of explicit isospin degrees of freedom to the system and of the isospin dependence to the mean field paves the way for future study on the nuclear symmetry energy as well as other isospin-dependent effects \cite{tsang2012constraints}.

\begin{figure}[t] 
   \centering
   \includegraphics[width=\textwidth]{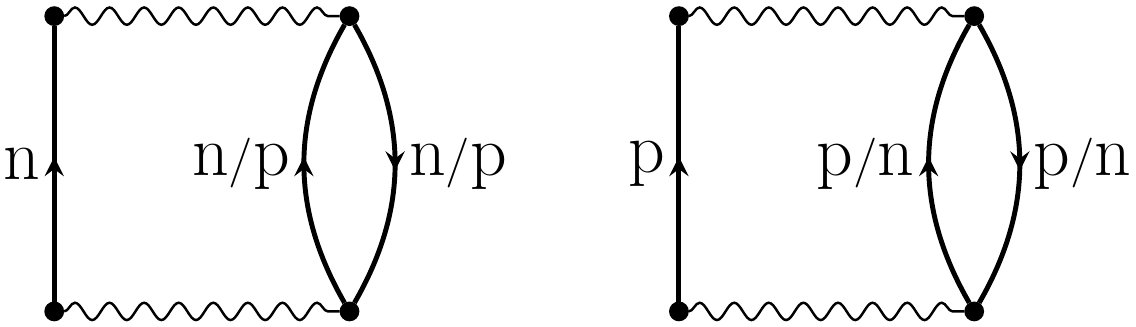}
   \caption{Second-order direct Born diagrams for neutron and proton self-energies $\Sigma_\textrm{n/p}$. Note that the particle species in the loop can be the same as or the opposite of the species of the propagator in the line to the left of the diagram.  In the present work, the interaction transfers no isospin within the Born diagram, making the scattering independent of isospin.}
   \label{diagrams}
\end{figure}

One of the advantages of the non-equilibrium Green's function approach, over the more conventional TDHF theory, lies in its ability to incorporate two-body interactions into the residual self-energies $\Sigma^\lessgtr$ in the correlation integrals. The incorporation of two-body interactions induces quantum correlations in the system, which have been lacking in both mean field approaches and semi-classical approaches at the explicit level. In the current approach, we have included in the self-energies the correlation contribution represented with the diagrams in Fig.~\ref{diagrams} by invoking the second-order Born approximation. Effectively, this direct diagram describes the scattering within $nn$, $pp$ and $np$ pairs with different lines representing Green's functions of the suitable species.  The corresponding self-energies $\Sigma^\lessgtr$ read
\begin{equation}
\Sigma^\lessgtr(p,t;p',t') = \int \frac{dp_1}{2\pi\hbar} \frac{dp_2}{2\pi\hbar} \, V(p-p_1) \, V(p'-p_2) \, G^\lessgtr(p_1,t;p_2,t') \, \Pi^\lessgtr(p-p_1,t;p'-p_2,t') \, ,
\end{equation}
where
\begin{equation}
\Pi^\lessgtr(p,t;p',t') = 2 \int \frac{dp_1}{2\pi \hbar} \frac{dp_2}{2\pi \hbar} \, G^\lessgtr(p_1,t;p_2,t') \, G^\gtrless(p_2-p',t';p_1-p,t) \, ,
\end{equation}
where $V(q)$ is the two-body potential with $q$ being the relative momentum and the factor of 2 accounts for spin degeneracy.

While some previous attempts were made to include residual two-body interactions \cite{rios2013towards,mahzoon2017correlations,*mahzoon2019nuclear}, the choice of the two-body potential $V(q)$ in those attempts might not have been optimal for the one-dimensional nuclear systems under consideration. Instead, to reflect the short-range nature of the residual nucleon-nucleon interactions and to mimic the effects of nucleon-nucleon scattering in the kinetic limit, we adopted a new form for the two-body potential:
\begin{equation}
V(q) = V_0 \, |q| \, \exp{\Big[-\Big(\frac{\eta q}{2\hbar}\Big)^2\Big]} \, ,
\label{eq:potential}
\end{equation}
where the values $V_0 = 205.491 \, \text{MeV}$ and $\eta = 0.57 \, \text{fm}$ have been chosen to establish a correspondence between the one-dimensional collision rates and the three-dimensional collision rates in symmetric nuclear matter in the semi-classical limit. 
Note that this potential yields no difference in the scattering between $np$ and $nn$ or $pp$ pairs.

\begin{figure}[t] 
   \centering
   \includegraphics[width=\textwidth]{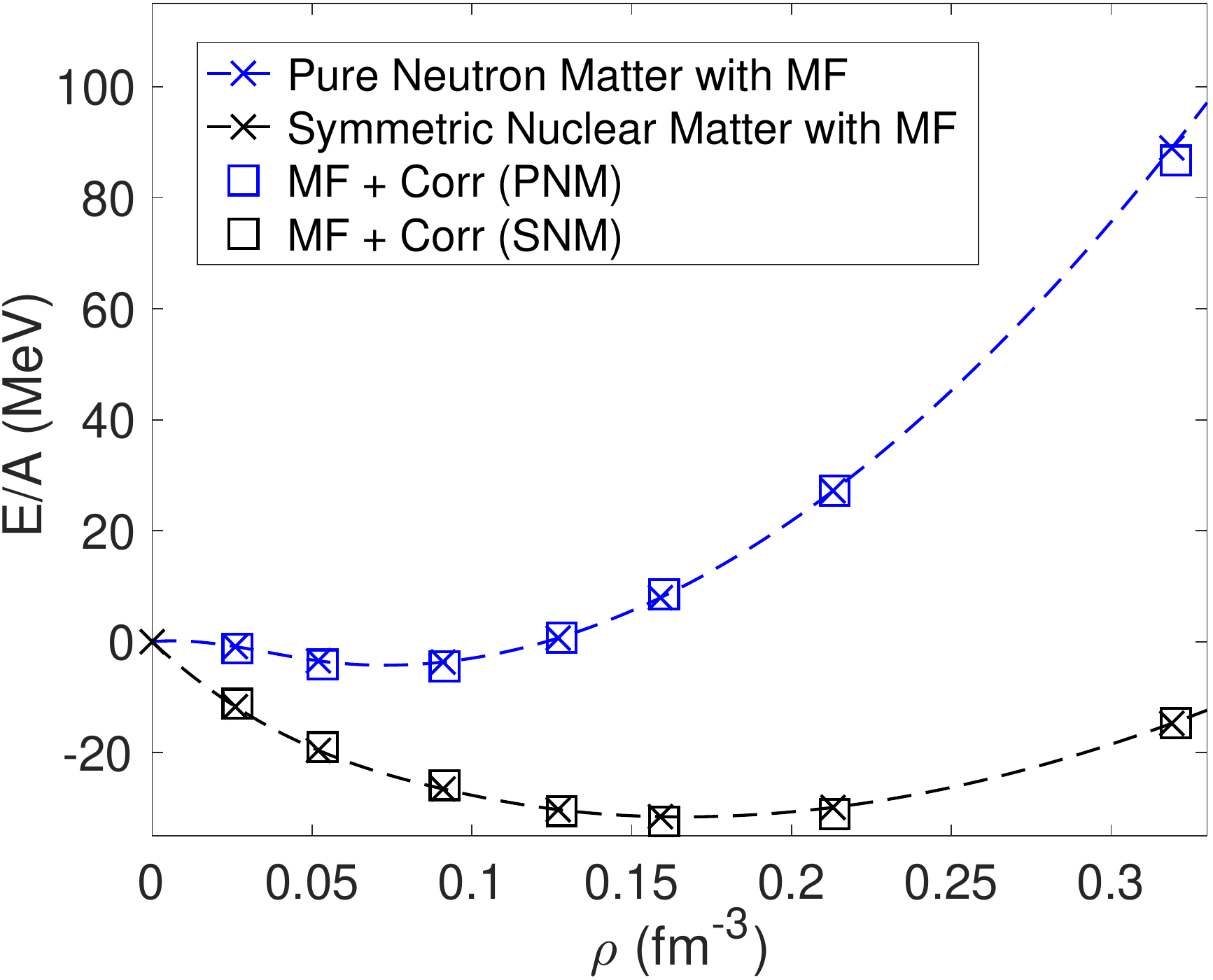}
   \caption{One-dimensional nuclear equation of state in the NGF approach.  Crosses represent the results obtained within pure mean-field approximation.  Dashed lines represent a cubic spline fit through the crosses.  Open boxes show results obtained within complete calculations with correlations, where the pure mean-field results were matched by adjusting the parameters in the auxiliary field in Eq.~\eqref{aux}.}
   \label{EOS}
\end{figure}

The residual interactions introduce an extra contribution termed $E_\textrm{corr}$ to the total energy and thus effectively alter the equation of state. This issue arises, as we have fitted the nuclear matter in the mean-field approximation to the nuclear matter equation of states in the first place, before introducing two-body interactions. Furthermore, in principle, the interactions entering the Hartree-Fock self-energy $\Sigma_\textrm{HF}$ and the residual self-energy $\Sigma^\lessgtr$ should be the same, which is beyond the naive separation of mean field dynamics and many-body dynamics employed in the current picture. Getting around the first issue, to get back to the previous satisfactory equation of state, we add an auxiliary parameterized contribution to the mean-field energy
\begin{equation}
\frac{E_\textrm{aux}}{A}(\rho,\delta) = A_1 \, \bigg(\frac{\rho}{\rho_0}\bigg)^{\tau_1}  \exp{\bigg(-b_1 \, \frac{\rho}{\rho_0}\bigg)} + A_2 \, \bigg(\frac{\rho}{\rho_0}\bigg)^{\tau_2}  \exp{\bigg(-b_2 \, \frac{\rho}{\rho_0}\bigg)} \, \delta^2 \, ,
\label{aux}
\end{equation}
and adjust parameters to reproduce the nuclear matter equation of state from the mean-field approximation in Eq. (\ref{hf}) as closely as possible.  The parameter values are summarized in Table \ref{Coeffs} and the results for the equation of state are illustrated in Fig.~\ref{EOS}.  The calculations are carried out for a uniform system enclosed in a box of size $L$, with $\Delta x = 0.57 \, \text{fm}$ and periodic boundary condition.

\begin{table}
\begin{ruledtabular}
\begin{tabular}{c c c c c c}
$A_1$ [MeV] & $\tau_1$ [MeV] & $b_1$ & $A_2$ [MeV] & $\tau_2$ [MeV] & $b_2$\\ \hline
39.33 & 1.157 & 1.192 & -14.66 & 1.533 & 0.70
\end{tabular}
\end{ruledtabular}
\caption{Parameters for the auxiliary field in Eq.~\eqref{aux}. \label{Coeffs}}
\end{table}

To give context to the above results, it should be pointed out that the discretization interval $\Delta x$ introduces a cut-off for particle momenta in the system, $p_\text{max} = h/(2 \Delta x)$.  That cut-off limits the integration over the momenta in the correlation self-energies, represented in Fig.~\ref{diagrams}, in addition to what the range parameter $\eta$ does.  Also the size $L$ of the periodic box determines the spacing for the mesh in momentum, $\Delta p = h/L$, and this impacts both kinetic and correlation contributions to the energy.  With this the auxiliary energy should have a dependence on $\Delta x$ and $L$, $\frac{E_\textrm{aux}}{A}= \frac{E_\textrm{aux}}{A}(\rho,\delta,\Delta x,L)$.  We treat $\Delta x$ as a parameter of our dynamical model, resigned to the fact that to get to analogous dynamic results we might need to adjust some other parameters, such as $V_0$ or $\eta$ in \eqref{eq:potential}, when changing $\Delta x$.
Here we choose $\Delta x = \eta$.  As to the dependence on $L$, we have carried out a number of calculations at different values of $L$ ranging from about $5$ fm to $25$ fm and different nucleon numbers corresponding to a wide range of density values of interest to us up to $2\rho_0$.  In these calculations we see variation less than $1 \, \text{MeV}$ in $E/A$ for uniform matter at fixed $\rho$, $\delta$ and $\Delta x$.

\section{Preparing finite correlated nuclear systems}\label{III}

Following the extensions to accommodate the isospin degrees of freedom and two-body residual interactions and the study of the equation of state for infinite nuclear matter, we turn to the preparation of  finite nuclear systems in one dimension. Obtaining a good approximation of the ground state is the starting point for performing simulations of nuclear collisions.

Within the full-fledged NGF approach, we continue to use adiabatic switching techniques~\cite{rios2011towards,rios2013towards,mahzoon2017correlations,*mahzoon2019nuclear,schluenzen2019ultrafast} to evolve the system towards an approximate ground state. At $t = 0$, the system is initialized in an uncorrelated harmonic oscillator ground state occupying the lowest $N/2$ neutron shells and $Z/2$ proton shells. The remaining spin degeneracy $g=2$ limits us to even-even nuclear systems. With the help of a switching function $f(t)$ varying smoothly from 0 to 1 over the switching period $[0, T_s]$, the total Hamiltonian $H(t)$ switches smoothly and slowly from the harmonic oscillator field $H_\textrm{HO}$ to the nuclear field $H_\textrm{nucl}$ according to $H(t) = [1-f(t)] \, H_\textrm{HO} + f(t) \, H_\textrm{nucl}$. The nuclear Hamiltonian $H_\textrm{nucl}$ contains both the mean-field component and the two-body interaction component. Since the second order diagrams in Fig.~\ref{diagrams} contain effectively interactions at two times, some extra care needs to be taken when adiabatically turning on the residual self-energy $\Sigma^\lessgtr$ which gets transformed according to:
\begin{equation}
\Sigma^\lessgtr(t,t') \rightarrow f(t) \, \Sigma^\lessgtr(t,t') \, f(t') \, .
\end{equation}
During the adiabatic switching process, the system evolves towards the nuclear ground state. The quality of the approximation to the ground state depends crucially on the switching time $T_s$, with a longer switching time yielding a better approximation. However, the entire preparation history enters the correlation integrals in Eq. (\ref{kbe}) in every single time step in the subsequent evolution. Too long a switching time can stall the numerical calculations quickly as the burden of history to be remembered grows.

The effects of a cooling friction during switching have been studied previously \cite{bulgac2013quantum,mahzoon2017correlations,*mahzoon2019nuclear}.  Imposing an additional friction-like potential $U^\textrm{fric}_{n/p}$ in the switching period can to a certain extent relax the need of an excessively long switching time.  The operational form that we employ for the potential is
\begin{equation}
U^\textrm{fric}_{n/p}(x,t) = f_0 \, \frac{\hbar}{\Delta t} \, \frac{\partial\rho_{n/p}(x,t)}{\partial t} \, \frac{t}{T_s} \exp{\bigg[- \frac{t}{T_s} \bigg]} \, ,
\label{eq:Ufric}
\end{equation}
where $f_0$ is the strength of the friction and $\Delta t$ is the time step size in the numerical simulations. While major desired dependencies have been factored out in Eq. \eqref{eq:Ufric}, the strength parameter $f_0$ remains fragile, with optimal values varying from system to system and from conditions to conditions.  In addition, even for optimal value any improvement in the switching with the inclusion of the cooling friction can be marginal.  In practice, extending the switching time still easily outperforms any delicately designed switching functions or cooling frictions, when a high-quality ground state is in need. For the time being, we simply implement the cooling friction parametrized in Eq. (\ref{eq:Ufric}) in the code with a relatively small strength $f_0 = 4-6$ fm$^4/$c, without claiming the optimality of the choice.

\begin{figure}[t] 
   \centering
   \includegraphics[width=\textwidth]{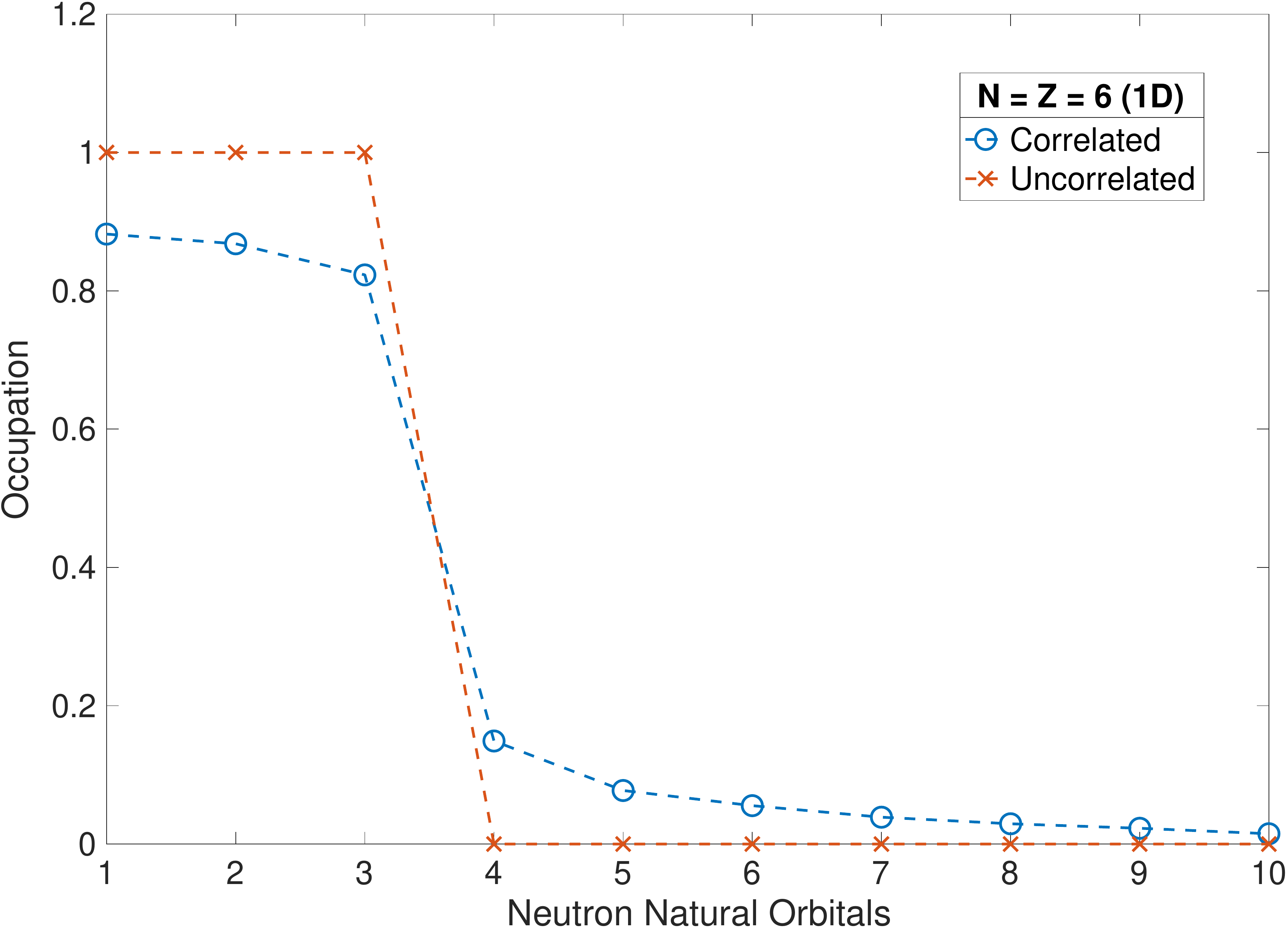}
   \caption{Occupation of the neutron natural orbitals. The open circles represent those of the correlated final state after the adiabatic evolution, and the crosses correspond to the uncorrelated initial state.  The lines serve to guide the eye.}
    \label{occupation}
\end{figure}

We now study the adiabatic evolution of a symmetric system with $N = Z = 6$. Initially, the system is comprised of 3 neutron shells and 3 proton shells in an external harmonic oscillator potential. The initial wave function can be written as an antisymmetrized product of the harmonic-oscillator single-particle states, and it is therefore an uncorrelated state. In Fig.~\ref{occupation}, we denote the occupation of the neutron natural orbitals, found by diagonalizing the one-body neutron density matrix, with red crosses. In the uncorrelated system, the natural orbitals coincide with the eigenstates of the single-particle harmonic oscillator hamiltonian $h_\textrm{HO}$, and the lowest three are fully occupied as expected. The case for protons is exactly the same due to the isospin symmetry present in the system and thus is not shown here. If the mean field were the only interactions switched-on in the system during the adiabatic evolution, the dynamics would be identical to that in the TDHF theory. Indeed, the NGF approach with mean-field approximation has been studied previously~\cite{rios2011towards,rios2013towards}. The final state would simply be uncorrelated, albeit as a product of complicated single-particle states. In the momentum basis $\{\phi_k \sim e^{ikx}\}$, however, even such an uncorrelated state becomes a complicated superposition of uncorrelated momentum states with all sorts of particle-hole excitations.  While in terms of the complicated states the density matrix can be diagonal, the density matrix acquires off-diagonal elements in the momentum representation.

The second-order Born diagram in Fig.~\ref{diagrams} allows for swapping of orbital pairs in an uncorrelated state due to the interaction and for mixing of that swapped state with the original, thus introducing a correlation impacting reinteraction.  Depletion in the probability of staying in the same uncorrelated state is compensated with a probability of populating other uncorrelated states.  In a system with weak nonuniformity, the density matrix becomes nearly diagonal in the momentum representation.  If the system is sparse or mixing is strongly suppressed by antisymmetrization, impact of a mixing persists over a long time. Eventually, a kinetic limit is reached where integrations in the residual self-energies can be carried out to arrive at energy-momentum conservation in a scattering process.  The peculiarity of the conservation in the two-particle scattering in one dimension is that the orbitals only either retain or interchange momenta.  For the species of the same type this leads to the persistence of the uncorrelated states, when composed out of near momentum eigenstates, and thus no apparent change compared to the mean-field dynamics.  However, because of the self-consistency in the diagram in Fig.~\ref{diagrams}, the admixtures of states with swapped orbitals pile up over time, leading to correlations between more than two orbitals even when residual interactions act between two particles only.  The scattering processes in the limit of energy-momentum conservation involve then more than two particles and more complicated changes occur in the uncorrelated states, potentially more similar to analogous processes in three dimensions than for the isolated two-particle scattering.  Also, for a more nonuniform system the different orbitals may share the same momentum, so even if the momenta are retained or swapped the occupations of uncorrelated states may change.  In~Fig.~\ref{occupation} we observe a significant depletion of the occupation for the low-lying states and increased occupation for the higher, in spite of the one-dimensional peculiarity of the on-shell two-body scattering and of the two-body scattering rate being coarsely set to be the same as for the three dimensions.  On the same note, even in three dimensions, when insisting on a quasiparticle picture for a finite system with energy conservation for single-particle energies, the departure of occupations from 1 and 0 would never occur.

If a particle is removed from or placed in a natural orbital from a stationary system with weak correlations, the system largely remains in a stationary state.  This can be quantified in terms of a spectral function with a unique single-particle energy, equal to the difference in the energies for the two stationary states.  Inspired by Fermi liquid theory, this independent quasiparticle picture is pushed to the extreme in the semiclassical transport theory for nuclear reactions where specific energies are attributed to quasiparticles with definite momenta~\cite{danielewicz2000determination}.  The spectral function naturally relates to NGF.  One characteristic of increasing correlation strength is that identification of natural orbitals becomes a challenge.  They may be always obtained by diagonalizing the density matrix but, as far as evolution of NGF in the relative time is concerned, the orbitals can be discussed only in a perturbative context. The~spectral function is not diagonal in any orbitals anymore, but can be near-diagonal and/or diagonal elements of the spectral function can be considered.

Given the above, in the correlated system we look at diagonal elements of the spectral function in momentum representation.  For particular species with particular spin projection, denoted with $\nu$ (also specifically used for neutrons in the shell-model notation),
the spectral function $S_\nu(p,E)$ is related to the retarded Green's function $G_\nu^+(p,E)$ with
\begin{equation}
S_\nu(p,E) = -\frac{1}{\pi} \, \textrm{Im} \, {G_\nu^+(p,E)} \, .
\end{equation}
The retarded Green's function $G_\nu^+(p,E)$ admits a Lehmann representation,
\begin{equation}
G_\nu^+(p,E) = \sum_\mathpzc{p} \frac{|\langle \Psi_\mathpzc{p}^{(N+1)}|\phi_\nu^\dagger(p)|\Psi_0^{(N)}\rangle|^2}{E-(E_\mathpzc{p}^{(N+1)}-E_0^{(N)})+i\eta} + \sum_\mathpzc{h} \frac{|\langle \Psi_\mathpzc{h}^{(N-1)}|\phi_\nu(p)|\Psi_0^{(N)}\rangle|^2}{E-(E_\mathpzc{h}^{(N-1)}-E_0^{(N)})+i\eta} \, ,
\end{equation}
where $\eta \rightarrow 0^{+}$, $\phi$ and $\phi^\dagger$ are annihilation and creation operators in momentum representation respectively, $\Psi_\mathpzc{p}^{(N+1)}$ are eigenstates of the total Hamiltonian with $N$+$1$ neutrons and $\Psi_\mathpzc{h}^{(N-1)}$ are eigenstates with $N$$-$$1$ neutrons.  The spectral function $S_\nu(p,E)$ picks out the poles of retarded Green's function, which are the excitation energies of adding or removing a neutron of momentum $p$ with respect to the $N$-neutron ground state $\Psi_0$. The strength of the excitation is regulated by the overlap function in the numerators in the retarded Green's function. On account of the completeness of the sets of states and commutation relation for the operators, the spectral function integrates over energy to~1.  This allows one to interpret $S_\nu(p,E)$ as the probability density for a particle $\nu$ with momentum $p$ to have an energy~$E$~\cite{bruus2004many}.

In a noninteracting Fermi gas the spectral function peaks at the kinetic energy of a particle with momentum $p$, i.e.~$S(p,E) = \delta[E-p^2\big/(2m)]$.  For a self-bound slab in a box, in the semiclassical limit behind the transport for reactions, two such peaks are expected, one at lower energy for slab interior and one at higher for the exterior, with relative intensity of the peaks representing the relative share of the size of the box for the slab interior and exterior.  In the Fermi liquid theory the spectral function of a correlated uniform system is often sought in the Lorentzian form: $S(p,E) \approx \frac{Z_p/(2\tau_p)}{(E-E_p)^2+[1/(2\tau_p)]^2} + S'(p,E)$, where $E_p$ is the energy of the quasi-particle with momentum $p$, $\tau_p$ is the lifetime, $Z_p$ is quasiparticle strength and $S'$ is contribution of incoherent background that ensures that the sum rule is satisfied~\cite{bruus2004many}.

\begin{figure}[t] 
   \centering
   \includegraphics[width=\textwidth]{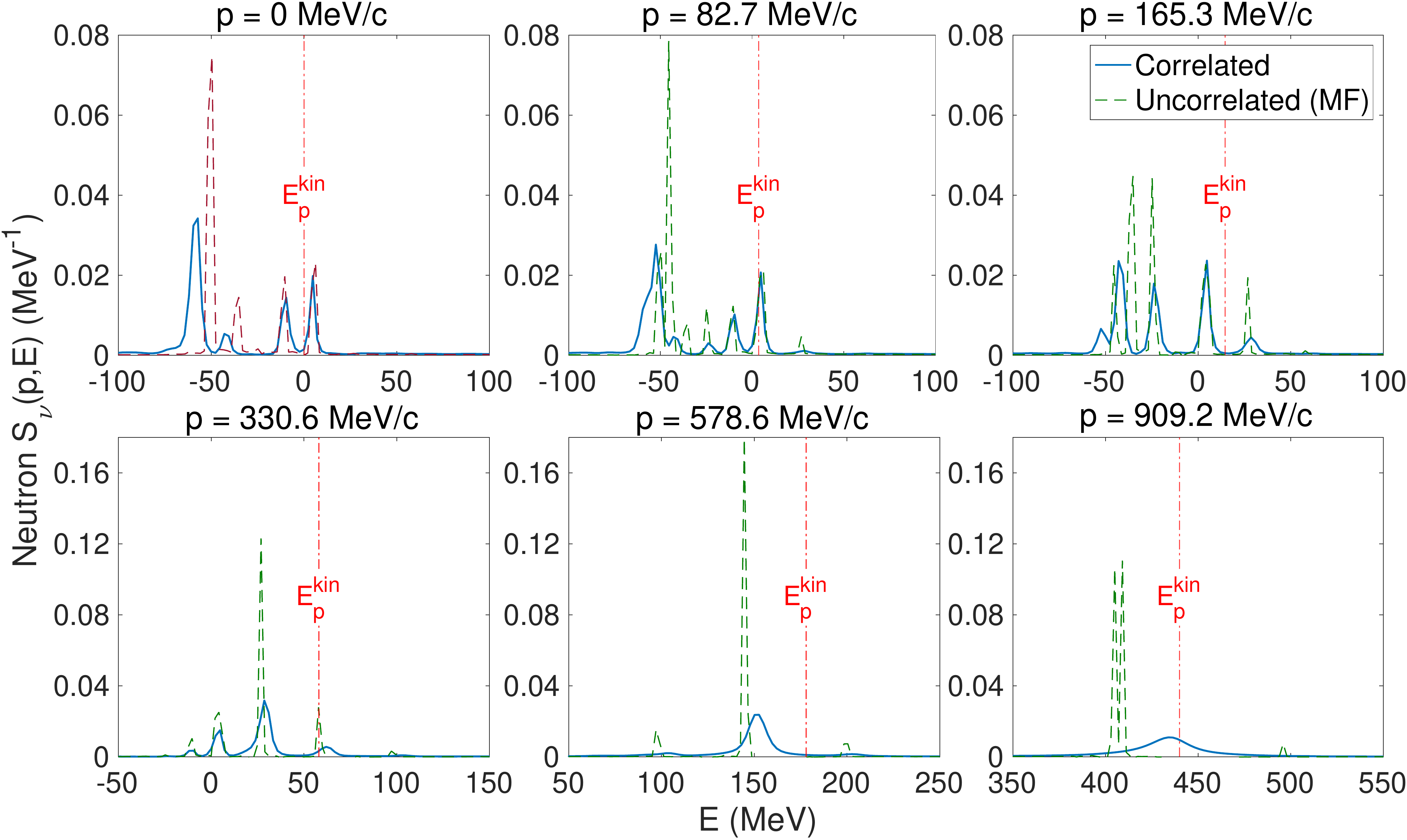}
   \caption{Neutron spectral functions vs energy for an $N=Z=6$ self-bound system in an $L=15 \, \text{fm}$ box at discretized momenta. The red dashed-dot line indicates the kinetic energy of a free neutron with momentum $p$. See the text for discussion.}
   \label{spectral}
\end{figure}

In Fig.~\ref{spectral}, we show the spectral functions for neutrons in a symmetric system with $N=Z=6$ in an $L=15 \, \text{fm}$ box, at six of the discrete momenta, in the mean-field calculation and in the full calculation with two-body residual interaction. The spectral functions are obtained by evolving the system over $300 \, \text{fm}/c$ and Fourier transforming the relative time difference $t_1\!-\!t_2$ of the Green's functions. For reference, we provide in the figure the location where the kinetic energy for a given momentum is. Multiple spikes are present in both calculations. For three of the lower momenta, strength is present at about $55 \, \text{MeV}$ down from the free-space kinetic energy, about the depth of the potential well in nuclear systems. The spikes in the mean-field case are sharper, and, for the three lowest momenta, a few of the peaks in the two different cases overlap almost perfectly, suggesting that the mean field alone is already responsible for a significant part of the fragmentation of natural orbitals in momentum representation. Notably in the limit of perfectly static infinite-time evolution, the mean-field peaks would turn into $\delta$-functions, meaning that any width observed for these peaks in Fig.~\ref{spectral} is due to our time domain methodology.  Inclusion of residual interactions smoothes out and spreads the peaks and moves some well beyond any resolution impact of our methodology.

The structures observed in Fig.~\ref{spectral} are outside of the reach of semiclassical transport theory.  Telling is the fact that to resolve the structures in the spectral function reasonably well, we had to evolve the system for over $300 \, \text{fm}/c$.  Key stages of reactions to which transport is applied commonly last a twentieth of that time~\cite{hong2014subthreshold,danielewicz2000determination}, meaning that energy structure will generally be poorly resolved and conservation in binary collisions will not hold~\cite{danielewicz1984quantumII}.

\begin{figure}[t] 
   \centering
   \includegraphics[width=\textwidth]{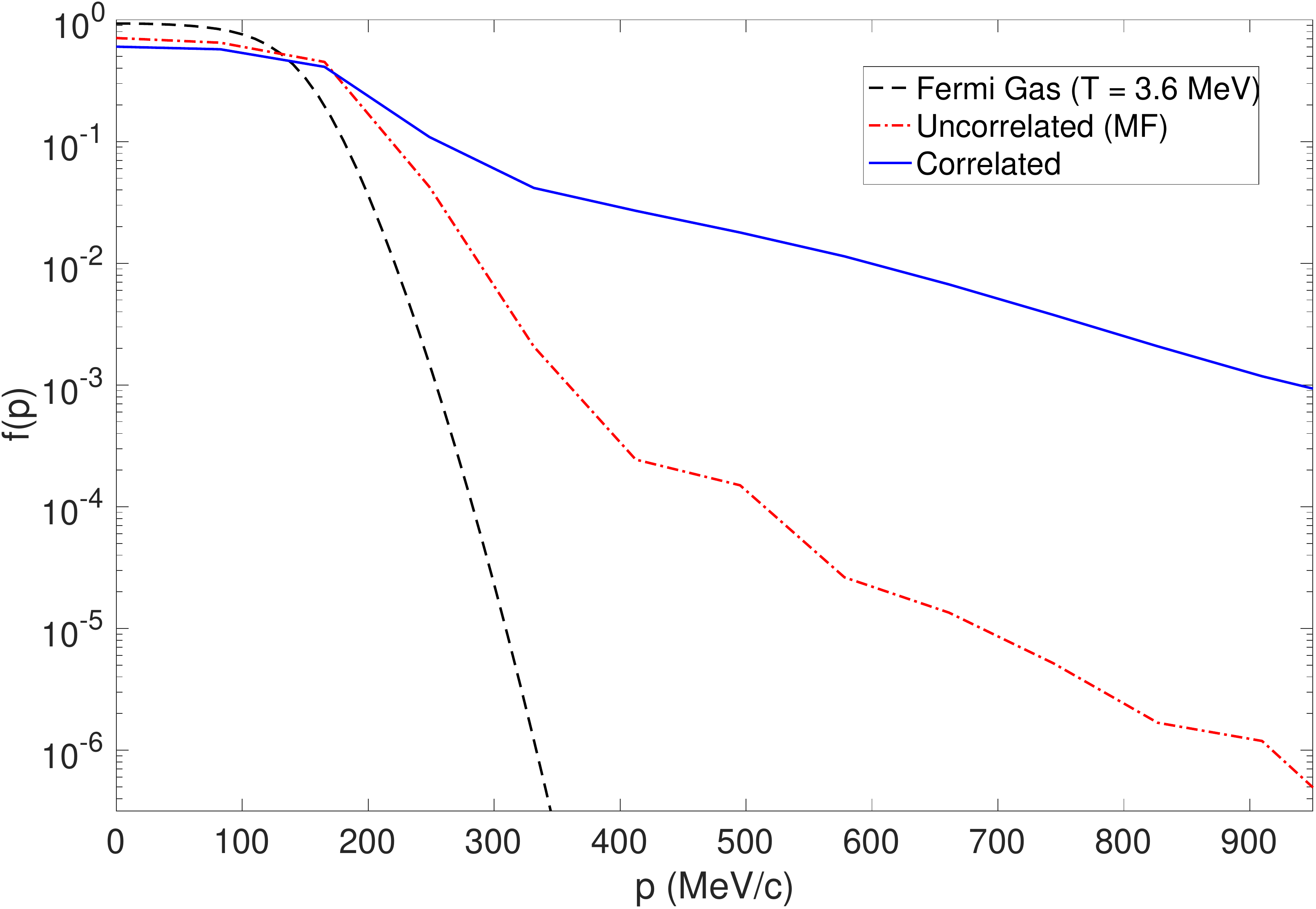}
   \caption{Momentum state occupation for neutrons in the same $N=Z=6$ ground-state system as in the preceding two figures and for a Fermi gas.  The dashed, dash-dotted and solid lines represent the Fermi gas at temperature $T = 3.6\,\text{MeV}$, the mean field system and the correlated system, respectively.}
   \label{momentum}
\end{figure}

In Fig.~\ref{momentum}, we show the momentum distribution for neutrons in the same $N=Z=6$ system we have been discussing and for a noninteracting Fermi gas at temperature $T = 3.6\,\text{MeV}$.  The high-momentum tails seen in the uncorrelated mean-field case and the case with short-range two-body interaction lack the features of a free Fermi gas at finite temperature, suggesting again that the underlying picture cannot be entirely captured in the Fermi liquid theory. In particular, the high-momentum ($> p_F^\textrm{1D} \approx$ 195 MeV/c) tail accounts for about~25\% of the neutrons in the correlated system, comparable to what is reported to be around~20\% in nuclei in nature~\cite{or2017, rios2014density, dickhoff2008many}, while the corresponding proportion in the uncorrelated system is just about 10\%. A significant number of the high-momentum nucleons are present due to the short-range two-body interaction.

\section{Exploring dissipation with isovector oscillation in a finite system}\label{IV}

To describe heavy-ion collisions at the intermediate energies ranging from tens of to hundreds of MeV/u, it is important to include dissipation of the relative energy in the center of mass frame in the dynamics, which is lacking in the conventional TDHF theory. However, as discussed in the preceding section, effects of the short-range two-body interactions included in the correlation integrals in the NGF approach can be interpreted as scattering and rearranging population of orbitals.  In fact, these integrals turn into collision integrals in the kinetic limit for the theory.

\begin{figure}[t] 
   \centering
   \includegraphics[width=\textwidth]{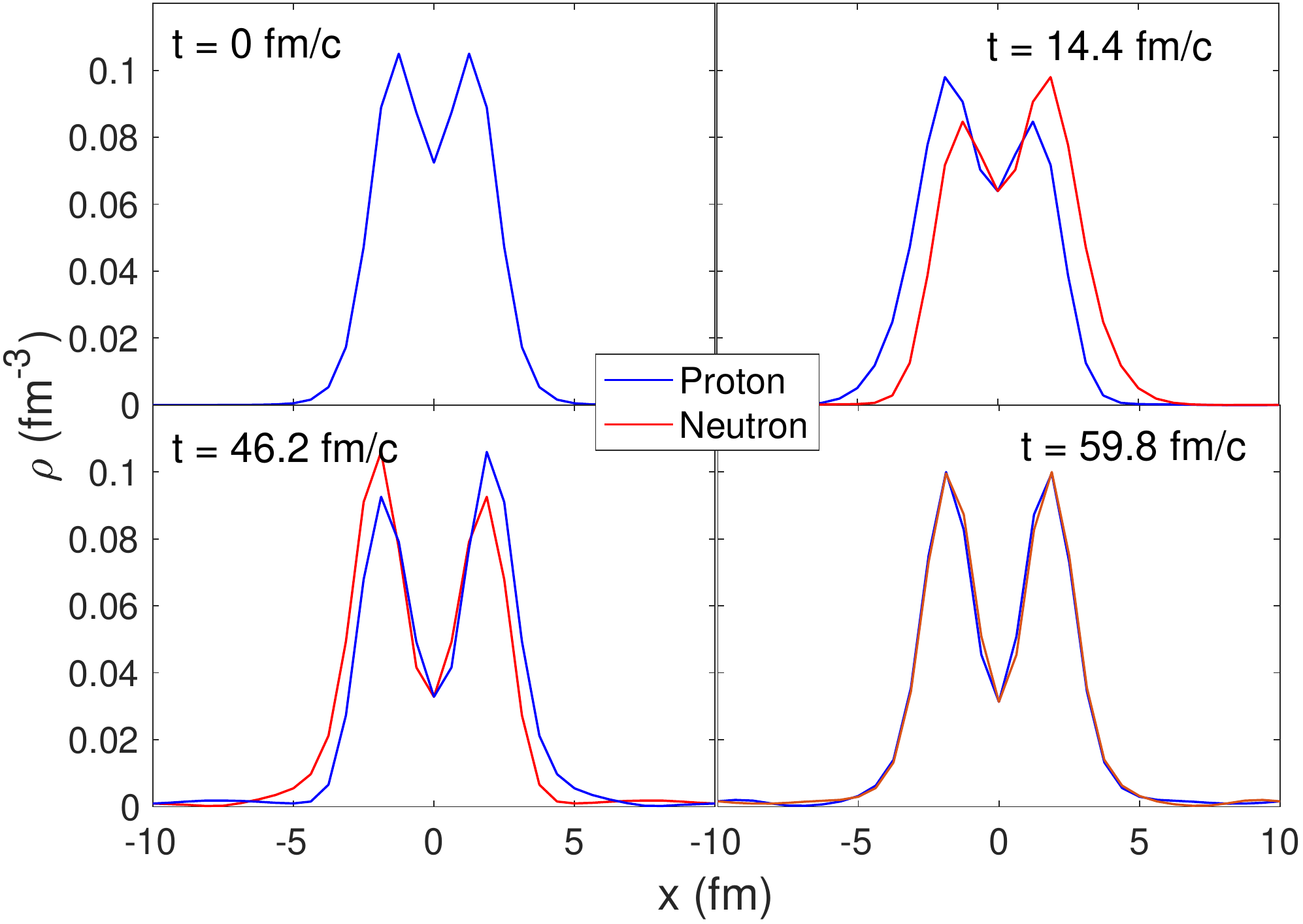}
   \caption{Isovector-like oscillation of the neutron slab and the proton slab with mean-field approximation. The red line represents the neutron density profile and the blue line represents the proton density profile.}
   \label{den_osc}
\end{figure}

\begin{figure}[t] 
   \centering
   \includegraphics[width=\textwidth]{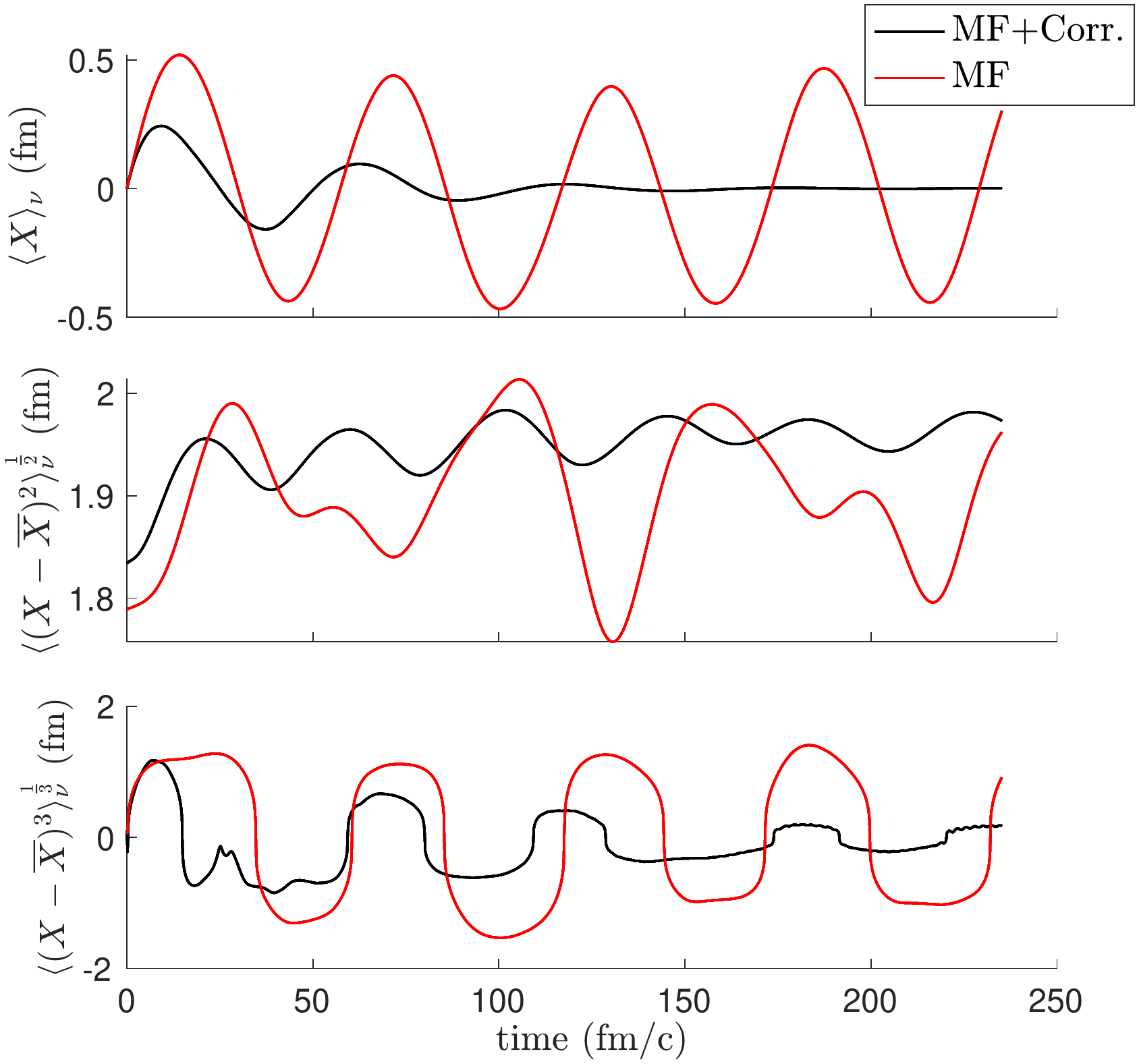}
   \caption{Evolution of the lowest three moments of the neutron slab. The red lines and the black lines correspond to calculations with mean field only and the full calculations, respectively.
}
   \label{osc}
\end{figure}

To explore the dissipative effects introduced by the short-range two-body interactions in the NGF approach, we simulate the isovector-like oscillation of neutrons and protons in a finite system. The simulations are performed in two different settings, one with only mean field dynamics, which is equivalent to the case in TDHF theory, and the other with both mean field dynamics and residual two-body interactions. For each setting, we prepare a symmetric nuclear system with $N=Z=4$ in its ground state at the center of a box, employing the adiabatic switching technique discussed before. The time is reset to $t = 0$ fm/c after completion of the adiabatic evolution. At $t = 0$ fm/c, we then perturb the neutron slab and the proton slab, which have been overlapping with each other perfectly. Specifically, we apply opposite jolts to the neutron and proton slabs, that take the form of Galilean boosts to the same-time lesser Green's functions $G^{<}_\textrm{n/p}$ at $t = 0$:
\begin{equation}
G^{\lessgtr,\,\textrm{after}}_\textrm{n/p} (x,0;x',0) = \exp{\Big(\pm i \frac{P x}{\hbar}\Big)} \, G^{\lessgtr,\,\textrm{before}}_\textrm{n/p} (x,0;x',0) \, \exp{\Big(\mp i \frac{P x'}{\hbar}\Big)},
\end{equation}
where $P$ is the small momentum boost of the order of $55 \, \text{MeV}/c$, equivalent to $1.6 \, \text{MeV}$ in kinetic energy per nucleon. After the jolts, the neutron and proton center of mass start to move away from one another, but as particles reach the edge of their optical potentials, they retreat and the centers of mass move back.  Fig.~\ref{den_osc} shows the evolution of the density profiles for neutrons and protons for the setting with mean field only.  The position of the center of mass for the neutron slab $\bar{X}$ can be seen, for both scenarios, in the top panel of Fig.~\ref{osc}.  When the residual interaction is present, the isovector oscillation is much more strongly damped.

On top of the isovector oscillation in Fig.~\ref{den_osc}, one may observe there deformations of the neutron slab and the proton slab with respect to their own center of mass. Since the slabs are not rigid, it is conceivable that one induces other forms of oscillation.  The additional changes may progress at the same or multiple of the isovector frequency, or at a completely different frequency.  For more insight we evaluate two more moments for a neutron slab and plot them as a function of time in the lower two panels of Fig.~\ref{osc}.  Specifically, the center panel displays the second moment about the center of mass of the neutron slab, $\langle (X-\bar{X})^2 \rangle^{\frac{1}{2}}$, describing the breathing mode for the neutron slab and also for the whole system after the isovector mode dies down.  Finally, the bottom panel displays the third moment about the center of mass of the neutron slab, $\langle (X-\bar{X})^3 \rangle^{\frac{1}{3}}$, representing skeweness of the slab around the center of mass.

The higher moments, quantifying behavior of the farther-out features of the density distribution, reach higher values in Fig.~\ref{osc} than does the center of mass position.  No matter what moment is considered the damping of its oscillations is much stronger for the dynamics with correlations than without.  In fact within the considered time interval hardly any damping is observed for the pure mean-field dynamics.  While the oscillations in the center of mass are fairly sinusoidal, indicating the dominance of a single frequency, this is less so for the higher moments.  The oscillations in the third moment are in phase with the those in the first, but are definitely anharmonic, suggesting that initialization of the Green's functions, with the same boost parameter at low densities as at high, might be too naive.  The second moments exhibit different periodicity than the odd moments, suggesting that an isoscalar breathing mode got excited and progresses at a different frequency than isovector.  Both the third and second moment damp away at a slower rate than the first moment, when the correlations are present. In the case of the third moment, the difference in damping compared to the first moment is subtle and can be due to the fact that the correlations weaken at the low densities that get tested by the high moment.  In the case of the second moment the damping is weak at longer time and we suspect that this is due to the peculiar nature of two-body scattering in the semiclassical limit in one dimension, where the orbitals either persist or interchange.  The interchange will shuffle the momentum between neutron and proton slabs, damping the isovector mode, but do nothing to the isoscalar mode once the isovector mode dies out.  The glitch in the third moment occurring at $25-30 \, \text{fm/c}$ is likely due to ripples in the density tails, produced by the boost, coming from both sides and colliding at the boundary of the periodic box.  Some early time transient behavior in the correlated dynamics is also due to the fact that the jolts are taken to be of a short duration compared to the relaxation of correlations.

\section{Galilean Covariance}\label{V}

The Kadanoff-Baym equations transform covariantly under Galilean transformation of the reference frame.  That means that Galilean transformed solutions from one frame should remain solutions of the equations in another frame.  Obviously there is no guarantee that this remains true in a numerical solution.  Discretization in space introduces a cut-off in momentum space that does not transform covariantly.  Practical violations of the covariance may impair studies of moving matter whether in collective oscillations or in collisions.  In~the context of the latter, we need to prepare slabs in their ground state, boost them and direct against each other.  Due to violation of covariance, the boost can potentially excite the slab and lead to its break-up ahead of any collision.

In this section, we explore impact of a boost on a slab in practice.  In the calculation we use a box of size $L = 15 \, \text{fm}$ with periodic boundary conditions.  We prepare a symmetric system with $N=Z=2$ at the center of the box as usual. We then boost the slab to provide it with a momentum per nucleon $P$.  When working with a periodic box, the momenta in representing the Green's functions are discretized with the spacing of $\Delta p = h/L$.  We simplify the boost by taking $P=2 \Delta p$, or wavevector $k=2 (2\pi/L) \simeq 0.838 \, \text{fm}^{-1}$, corresponding to boost velocity $V=P/m=0.176 \, c$.  With this the Green's functions can be shifted over momentum mesh without interpolations.

The correlation integrals in the Kadanoff-Baym equation involve the integration of the entire history of the system, rendering the dynamics non-Markovian, in sharp contrast to both semi-classical transport theories and the TDHF theory. If the boost were applied to the end of the adiabatic evolution, we would be boosting the single-particle characteristics, but not two-particle correlations.  We can afford it to a degree when exciting isovector oscillations, risking some heating of the system in addition to that occurring anyway due to dissipation of the collective oscillation.  The local Galilean boost operator $\mathcal{T}(x,t;V)$, that transforms a Green's function from a frame of a still system with positions in terms of coordinate $x$ to one in terms of $x'=x+Vt$, according to
\begin{equation}
G^{\lessgtr}_\textrm{moving}(x_1',t_1;x_2',t_2) = \mathcal{T}(x_1,t_1;V) \, G^{\lessgtr}_\textrm{still}(x_1,t_1;x_2,t_2) \, \mathcal{T}^*(x_2,t_2;V) \, ,
\end{equation}
is
\begin{equation}
\mathcal{T}(x,t;V) = \exp{\Big[\frac{iP}{\hbar} (x+Vt/2) \Big] } \, .
\end{equation}
In momentum representation this yields
\begin{equation}
G^{\lessgtr}_\textrm{moving}(p_1,t_1;p_2,t_2) = \textrm{e}^{-\frac{iV t_1}{\hbar}(p_1-P/2)} \,
G^{\lessgtr}_\textrm{still}(p_1-P,t_1;p_2-P,t_2) \, \textrm{e}^{\frac{iV t_2}{\hbar}(p_2-P/2)} \, .
\end{equation}
When starting investigation of a boosted slab at $t_0=0$, the transformation is applied to functions where one of the arguments coincides with $t_0$ and the other either coincides with or precedes $t_0$.

\begin{figure}[t] 
   \centering
   \includegraphics[width=\textwidth]{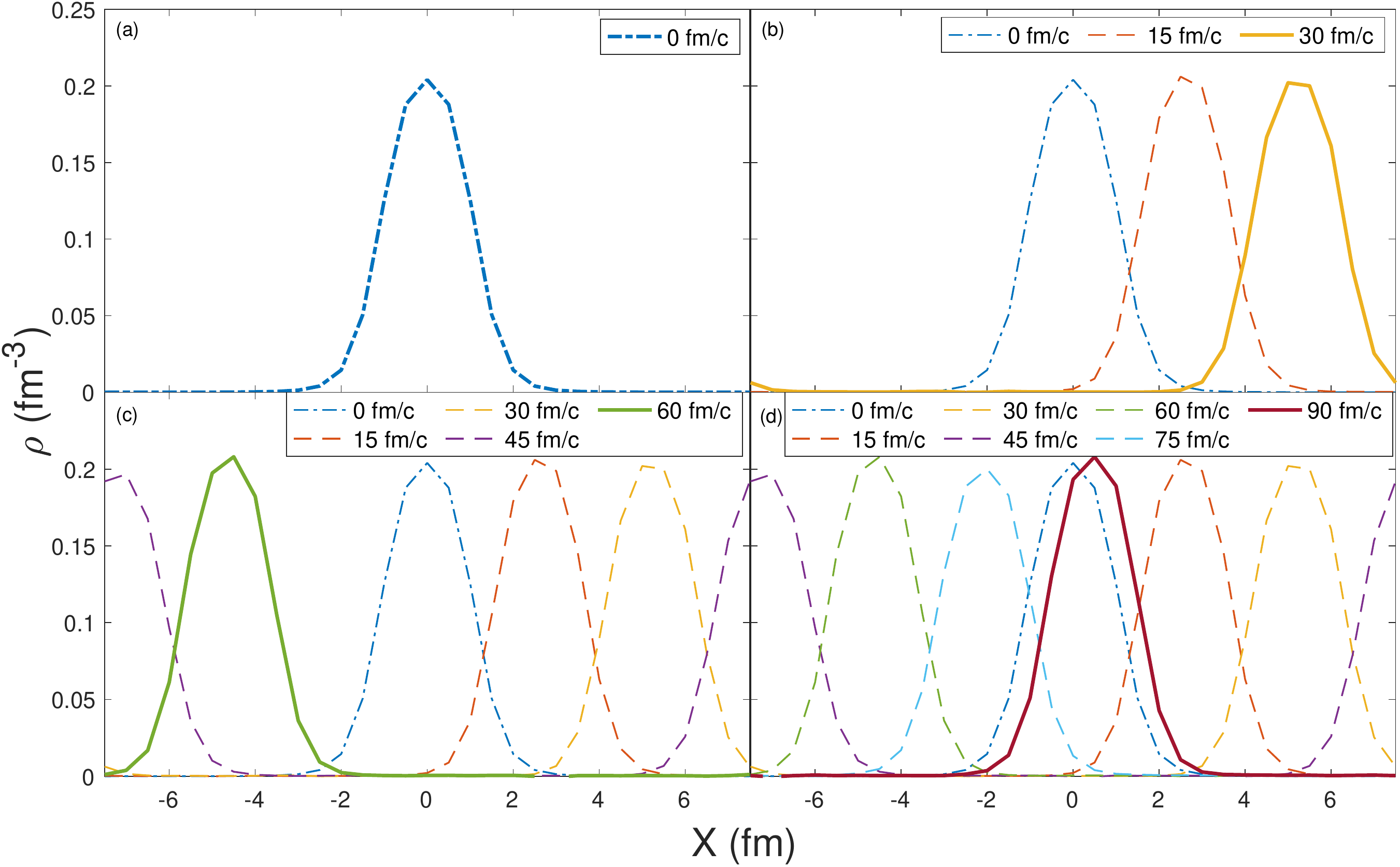}
   \caption{Evolution of density for a correlated $N=Z=2$ slab boosted with velocity $V=0.176 \, c$.  The initial density is illustrated with a dash-dotted line in every panel.  The panels (b)-(d) show evolution advanced subsequently over $30 \, \text{fm}/c$, with the latest profile illustrated with a solid line.  Intermediate profiles at every $15 \, \text{fm}/c$ are illustrated with dashed lines.}
   \label{translation}
\end{figure}

After applying the boost transformation in momentum, we evolve the system according to the Kadanoff-Baym equations in the standard manner.  The evolution of the density profile for a boosted slab is illustrated in Fig.~\ref{translation}.  Specifically in panel (a) we show the initial profile with a dash-dot line.  In the subsequent panels (b)-(d), we subsequently advance the evolution by $30 \, \text{fm}/c$ in each, displaying the latest profile with a solid line and intermediate profiles, at $15 \, \text{fm}/c$ intervals, with dashed lines.  It is apparent that the slab moves at a~uniform pace.  Some small variations in the profile are seen, potentially related to spatial discretization, but these are not significant enough to be of concern in exploring slab collisions.  With this, we demonstrate covariance of the NGF approach adequate enough for future studies of the collisions.

\section{Summary and future prospects}\label{VI}
In this paper, we have discussed the application of the NGF theory to the description of the dynamics of one-dimensional correlated nuclear systems in different scenarios. The~preparation of finite nuclear systems is relevant for the initialization of projectile and target in simulating nuclear collisions. The dissipation arising from the short-range two-body interactions, demonstrated in the isovector-like slab oscillation, can give rise to the stopping effects in collisions. One should bear in mind that the dissipation in one dimension simulated in the current numerical model may not have captured the entirety of the actual dissipation in three dimensions. In a three-dimensional description, relative longitudinal momentum can be effectively redirected into the transverse directions, presumably enhancing the dissipation. In the current one-dimensional description, dissipation effects are still largely confined to the internal excitation of the systems in terms of exchanging momentum. The translation of a stable slab partially mimics the situation of two nuclei approaching each other before making contact, but we do acknowledge that boosting two stable slabs in the opposite directions requires more careful design and is thus out of the scope of this paper.

We shall also briefly touch on the limitations and possible future extensions of the current NGF model. The tremendous amount of CPU hours and memories needed to solve the Kadanoff-Baym equation numerically renders direct simulations in higher dimension computationally infeasible. In fact, even in the current model, we needed to look for techniques to reduce the computational cost and the calculations have only been possible on high-performance computer clusters. We have used OpenMP~\cite{chandra2001parallel} to parallelize the code so that independent loop iterations can be carried out simultaneously on multiple cores. We also made the observation that, at any given moment $t$, in propagating the Green's functions $G^{<}(\cdot,t',\cdot,t)$ and $G^{>}(\cdot,t,\cdot,t')$ with $t\ge t'$, only the history correlated with the current is relevant, i.e., quantities with time arguments $(t,t')$ or $(t',t)$. This allows for the use of arrays with just one dimension in time for $t'$, which can be overwritten during the propagation of $t$, in sharp contrast to the naive way of storing quantities in both of the temporal dimensions. The~optimization made so far makes no compromise on the exactness of the numerical solutions in the stated framework. Further reduction in time or memory is possible under the assumption that elements far off the diagonal are negligible.  In fact, in Ref.~\cite{rios2011towards} it was demonstrated that far off-diagonal spatial elements of Green's functions may be dropped with impunity and one may hope to be able to introduce a finite memory time limiting extent in time.  These could reduce the time and memory cost of computations.

Regarding extension of the scope, the current separate parametrizations of mean field and residual interactions represents some limitation.  Progress towards consistency between the interactions could be reached by solving the $T$-matrix equation~\cite{danielewicz1984quantum} with an interaction more representative of a bare interaction.  With a separable approximation to the latter, the computational effort would not be much more serious than the current.  At low energies, excitations of chaotic collective modes can compete with nucleon-nucleon collisions in equilibrating larger colliding systems.  Accounting for such excitations can be achieved with a ring approximation to the self energy, again requiring quite similar computational effort to the current.  Beyond the short-range correlations, extensions can be conjectured to the current NGF model to treat pairing correlations upon differentiation of the spin degrees of freedom.  Paring effects have been incorporated into collisions in a three-dimensional TDHF code through the initialization of Bardeen-Cooper-Schrieffer (BCS) ground states \cite{maruhn2014tdhf}, with the caveat that the paring amplitudes are kept fixed during the evolution. A~more proper way to treat paring correlations in time-dependent processes is to solve the TDHF-Bogoliubov (TDHFB) equation~\cite{cambiaggio1984time,avez2008pairing,ebata2010canonical}. It is unclear how paring can enter the current picture of the NGF theory. Still, following the footprints of the TDHFB theory, one may try including a~time-diagonal paring-Green's functions of the form $\langle \phi(x_1,t) \, \phi(x_2,t)\rangle$, analogous to the abnormal density or the pairing tensor~$\kappa(t)$ in the TDHFB theory. A paring term~$\Delta$ shall enter the Hamiltonian when the Kadanoff-Baym equations for the normal Green's functions are solved. An additional equation for the time-diagonal paring-Green's functions void of the short-range interactions, mimicking that for the paring tensor~\cite{ebata2010canonical}, also needs to be solved simultaneously. Within such a naive conjecture, violation of total particle number and total energy might arise though.

In the end, we conclude that this work extends the application of the NFG theory in nuclear physics beyond the mean field approximation and lays the ground for the future study of nuclear collisions, taking us one step closer to a fully quantum-mechanical and realistic transport model. A multitude of extensions and optimization can also be considered and implemented for the improvement of this model.

\begin{acknowledgments}
The authors are thankful to Scott Pratt for fruitful discussions. This work
was supported by the U.S.\ National Science Foundation under Grant PHY-1520971, the U.S.\ Department of
Energy Office of Science under Grant DE-SC0019209 and the UK Science and Technology Facilities Council (STFC) through grant ST/P005314/1.
\end{acknowledgments}

\bibliography{NGFCorr}

\end{document}